\documentclass[prd,twocolumn,floats,floatfix,nofootinbib,tightenlines,superscriptaddress]{revtex4}
\usepackage{graphicx}
\usepackage{epstopdf}

\newcommand{\gev}{{\ensuremath\rm GeV}}
\newcommand{\tev}{{\ensuremath\rm TeV}}

\newcommand{\fb}{{\ensuremath\rm fb}}

\newcommand{\ifb}{{\ensuremath\rm fb^{-1} }}
\newcommand{\lum}{{30 } }

\newcommand{\h}{{\it H}}
\newcommand{\nness}{{\tau_N^{j}}}
\newcommand{\diness}{{\tau_2^{j}}}
\newcommand{\nnessCut}{{0.08}}
\newcommand{\signalsigma}{{2 }}
\newcommand{\mWidth}{{10}}

\begin{document}

\title{Rest Frame Subjet Algorithm with SISCone Jet for Fully Hadronic Decaying Higgs Search}
\author{Ji-Hun Kim}
\affiliation{FPRD and Department of Physics, Seoul National University, Seoul, 151-747, Korea}
\affiliation{Theory Division, CERN, CH-1211 Geneva 23, Switzerland}

\begin{abstract}
The rest frame subjet algorithm
is introduced to define the subjets for the SISCone jet;
with this algorithm,
an infrared and collinear safe jet shape observable N-subjettiness, $\nness$, is defined
to discriminate the fat jet, from a highly boosted color singlet particle decaying to N partons, 
from the QCD jet.
Using rest frame subjets and $\diness$ 
on dijets from highly boosted $\h / W / Z$ bosons through $pp \to \h W$, $\h Z$ with $m_H =120 \gev$,
we found that  
statistical significance of the signal, from the fully hadronic channels, is about  $\signalsigma \sigma$ 
for $14 \tev$ collisions with $\mathcal{L} \sim \lum \ifb$.

\end{abstract}

\maketitle

Current electroweak precision fits of the standard model(SM) prefer a light Higgs boson\cite{Alcaraz:2009jr};
experimental bounds from LEP\cite{Alcaraz:2009jr} and Tevatron\cite{:2010ar} 
suggest the SM Higgs mass, $m_H $, is about $120 \gev$.
However, the SM Higgs boson $m_H \lesssim 135 \gev$
has been considered hard to discover
because it dominantly decays to a b-quark pair.
Unlike the signals from leptonic decay modes, 
those from hadronic decay channels undergo overwhelming QCD background processes.
Finding the Higgs with $m_{\h} \lesssim135 \gev$
requires combining many possible decay channels  such as 
$\h \to \gamma \gamma$, $WW$, $b\bar{b}$ and $\tau\tau$.

Recently, there has been a study on the jet substructure via $pp \to W\h, Z\h$  
where $\h$ is highly boosted to form a single fat jet, and $W/Z$ decays leptonically\cite{Butterworth:2008iy}.
According to ATLAS detector simulation\cite{ATL-PHYS-PUB-2009-088}, 
it is expected to also be a promising channel for the Higgs search.
Other subjet techniques are also proposed for reconstructing the mass peak of heavy particles
\cite{Plehn:2009rk, Ellis:2009me, Soper:2010xk}
and reducing the QCD contamination effects\cite{ Krohn:2009th}.
Jet radius  also plays an important role for the mass peak reconstruction\cite{Cacciari:2008gd}, 
and several schemes for optimizing jet radius to improve the invariant mass distribution 
are proposed \cite{Krohn:2009zg, Soyez:2010rg}.

Beside jet substructure, ref. \cite{Almeida:2010pa} suggested `template overlap' method 
to match jet energy flow with the partons directly.
The property of color structure of the Higgs decay has also been studied 
for searching new particle via the double diffractive process \cite{Albrow:2008pn} 
and the Higgs decaying to the two b-tagged jets \cite{Gallicchio:2010sw}.


To search heavy particles, however, those techniques
usually require signal processes to involve
additional unusual signatures, such as lepton or missing-$p_T$,
to suppress QCD background;
fully hadronic decay channels, such as 
Higgs production associated with $W / Z$ boson, 
$pp \to\h W, \h Z$,
are still considered too hard to be used for the heavy particle study.
Usually, however, cross-sections of hadronic decay channels are, at least, few times larger 
than (semi-)leptonic decay channels;
utilizing them will improve, for instance, the Higgs discovery potential.

In this paper, to identify boosted color singlet particles,
we introduce the subjet definition for SISCone jet \cite{Salam:2007xv} 
and an infrared and collinear safe jet shape observable, $\nness$.
With $\diness$, as an illustration, 
we investigate the statistical significance of the signals from fully hadronic decay channels of light SM Higgs.

SISCone jet algorithm 
is one of the best algorithms for reconstructing mass peaks,
\textit{i.e.} it has low sensitivity to underlying event(UE) contamination \cite{Cacciari:2008gd, Sapeta:2010uk}.
In many cases including light SM Higgs searches, however,
such mass peaks are ruined by huge QCD background contributions.
To suppress the background,
the underlying structure of the jet is essential, 
but SISCone jet algorithm misses the informations.
Moreover, the jet substructure is also required for the boosted Higgs jet tagging.
The boosted Higgs jets contain not a single, but two secondary vertices 
originating from two b-quarks. 
The identification of the two b-quarks is crucial 
for separating the signal from the large background;
since gluon splitting into b-quarks is major source of QCD background,
the directions of the subjets can be used to reduce the background 
\cite{ATL-PHYS-PUB-2009-088}.
Previously, such two $b$ subjets tagging algorithms 
are devised for sequential recombination jet algorithms; for example, Cambridge/Aachen(C/A) algorithm\cite{ca_algo}.
With a subjet definition for SISCone jet, however,
those two $b$ subjets tagging algorithms can also be employed with SISCone jet, too.
These reasons have motivated us to devise a subjet algorithm using the SISCone jet algorithm
for finding a boosted particle while reducing QCD background.

We considered several possible ways to define the subjet of the SISCone jet.
One is, for a given jet, clustering its constituents with a sequential recombination jet algorithm
such as C/A with mass drop and filtering(MD-F) techniques \cite{Butterworth:2008iy}.
However, sequential recombination jet algorithms require a larger initial jet radius
than SISCone jet algorithm to catch perturbative radiation of the particle;
applying C/A MD-F to the SISCone jet is less efficient than using C/A MD-F solely.
Other one is clustering the constituents of the jet using SISCone algorithm with a smaller jet radius,
\textit{i.e.} applying the jet trimming algorithm\cite{Krohn:2009th}.
As the Higgs is boosted highly, however, the cones which define subjets are likely to overlap.
Although SISCone algorithm merges the overlapping cones or splits them according to the overlap threshold parameter,
we found that this subjet definition is inefficient, too.
Last method we considered is to define the subjet in the `rest frame' of the jet.

The fat jet from a boosted color singlet particle is, a good approximation, a closed system,
\textit{i.e.} it does not interact with the rest of the system when it hadronizes.
Thus, the rest frame of the particle is identical to the CM frame of its decay products.
In contrast, 
a jet from a colored particle is not a closed system.
Although colored partons are almost free at high energy due to asymptotic freedom,
they must hadronize and the hadronization process involves exchanging the four momentum.
Because the four momentum of the jet is different from that of the hard parton,
the rest frame of a colored particle has ambiguities.

\begin{figure}
\includegraphics[width=0.45\textwidth]{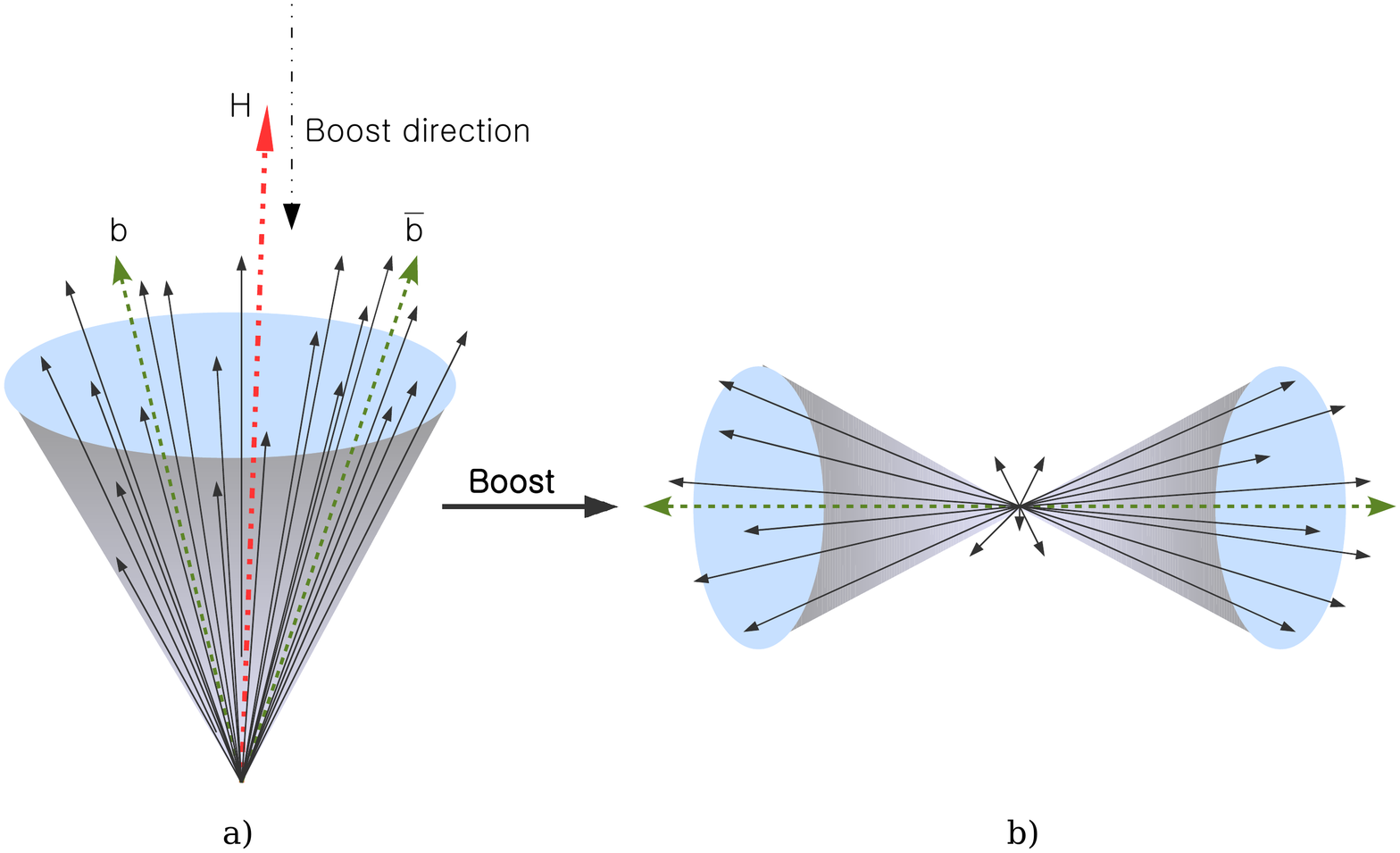}
\caption{Illustration of the jet and the rest frame subjets. 
a) The lab frame, b) The jet rest frame.}
\label{JetRestFrame}
\end{figure}

We define the jet rest frame as a frame
where the four momentum of the jet equals $p_\mu^{rest} \equiv (m^{jet}_{inv}, 0, 0, 0)$.
\footnote{We found that the similar concept for the top tagging exists in \cite{restTop}.}
A jet is consisted of its constituent particles.
The distribution of constituent particles of a fat jet, in their center of mass frame, is nearly identical to
those of the color singlet particle which is produced at rest.
Thus, we recluster them in the jet rest frame, 
\textit{i.e}. using their four momenta at the rest frame of the jet.
For a given jet, the `rest frame subjets' are defined as these reclustered jets. 
This procedure is illustrated in Fig. \ref{JetRestFrame}. 


UE affects the jet mass and, thus, the jet rest frame.
Consider a fat jet of $p_\mu = \gamma m (1, \vec{\beta})$, defined by a cone with angle, $\theta_c$.
In the jet rest frame, the energy of massless UE particles  
can be, at most, $\gamma (1 - \beta \cos \theta_c)$ times larger than its energy in the lab frame.
For UE particles with  $E < 1 \gev$, inside the SISCone jet with a jet radius 0.8 and $\gamma \sim 10$, 
the factor is about 3; soft massless objects in the lab frame remain soft in the jet rest frame.
Moreover, they are collimated to the boost axis, $\cos \theta \sim 0.97$,
and unlikely to included the leading rest frame subjets;
to be identified as a single fat jet, the angles between the axis and the leading rest frame subjets
are usually more large.
Thus, the rest frame subjet algorithm is infrared and collinear safe
if an infrared and collinear safe jet algorithm is used for the rest frame subjet clustering,
and the leading rest frame subjets of fat jets do not much affected by UE particles.

In the jet rest frame, we can treat the constituent particles of the jet
as the final state particles of a fictitious event, because it is the center of mass frame for the particles.
For example, in the rest frame of the Higgs jet, 
the constituent particles look like a dijet event from the Higgs produced at rest.
To have such a shape, the QCD jet should radiate only one hard parton which is improbable.
Since the gluon and quark jets are not closed systems,
their shape in the rest frame 
does not correspond to any physical state, 
and is more likely to be irregular.
Thus, by checking whether a shape of a jet in the rest frame looks like `N-jet' event,
\textit{i.e.} by analyzing whether the jet has N `rest frame subjets', 
we can discriminate between the fat jet, from the boosted color singlet particle decaying to N partons,
and the QCD jet. 

To select jets which have N rest frame subjets,
we employ the N-jettiness \cite{Stewart:2010tn}.
A global event shape ``N-jettiness'', $\tau_N$, 
is devised to filter out events which have additional undesired jets beyond the required N jets.
Treating the constituent particles of the jet as final state particles of a fictitious event,
we can apply $\tau_N$ to the jet, \textit{i.e.} calculate ``N-subjettiness'', 
to determine whether the jet has N rest frame subjets.
$\tau_N$ vanishes in the limit of exactly N infinitely narrow jets;
we expect the fat jets tend to result smaller $\nness$ than that of the QCD jet.

Except for it is defined by 
constituent particles of a jet instead whole final state particles of a event,
the definition of $\nness$ is identical to that of $\tau_N$.
In the jet rest frame, we tag N most energetic subjets, and define $\nness$ as :
\begin{equation}
\nness\equiv \frac{2}{ (m^{jet}_{inv})^2} \sum_{k \in J}  
  \min \{q_1 \cdot p_k, q_2\cdot p_k, \dots,  q_N\cdot p_k\} 
\end{equation}
where $p_k$ is four momenta of the constituent particles of the jet, J,
and $q_i$ is four momenta of the $N$ energetic subjets.
Definitions of $\tau_N$ and the rest frame subjet are infrared and collinear safe;
and, thus,  $\nness$ is also an infrared  and collinear safe observable.

\begin{figure}
\includegraphics[width=0.45\textwidth]{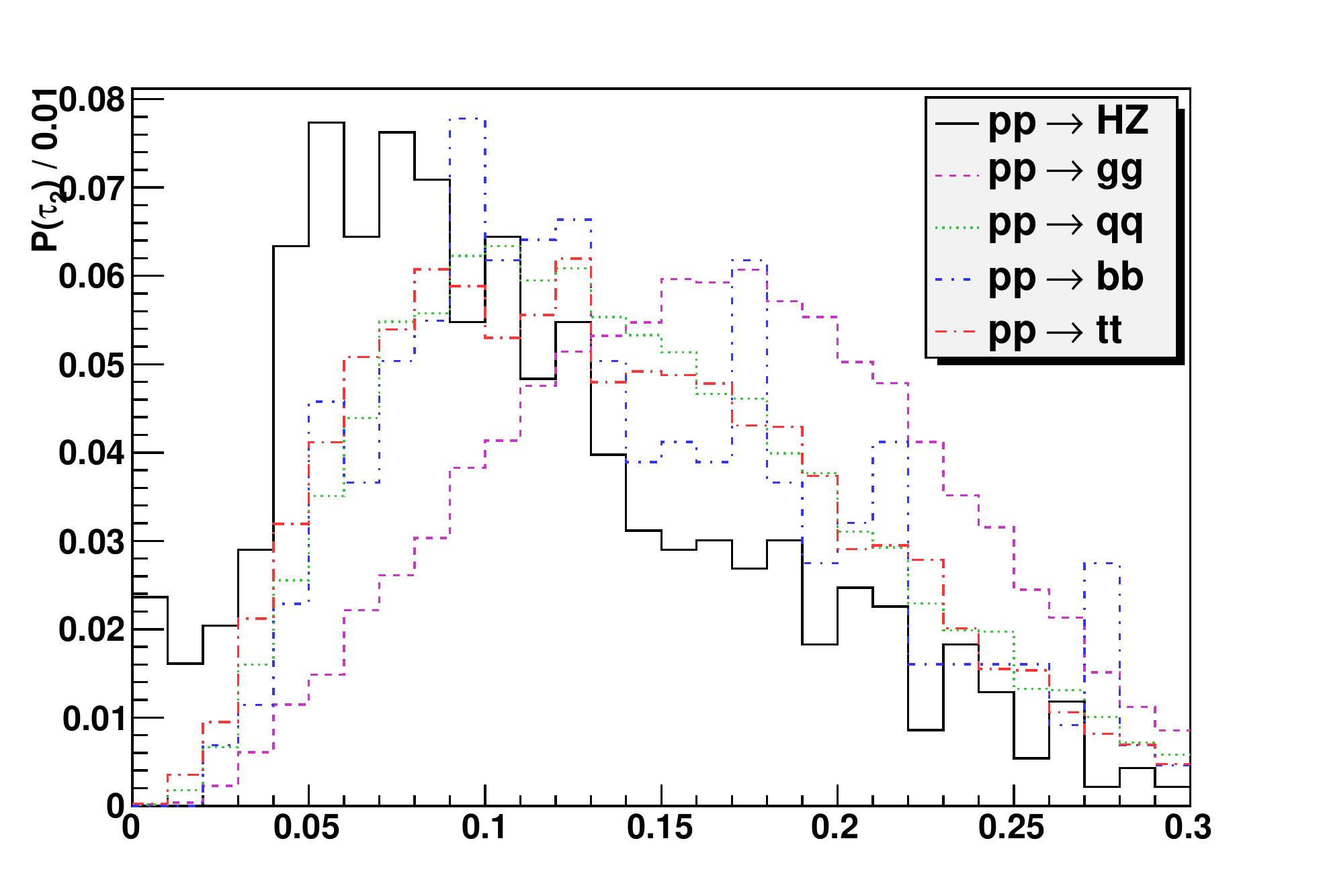}
\caption{$\diness$ distributions of jets with 
$ 110 \gev \le m^{jet}_{inv}  \le 130 \gev$, $p_T > 200 \gev$, $|y| < 2.5$.}
\label{nJetti}
\end{figure}

$\nness$ is calculated through following steps.
Given a hard jet $J$ with $p^J_\mu$
and its constituent particles $\lbrace i| i \in J\rbrace$ :
\begin{enumerate}
\item Define the boost vector, $\vec{\beta}_J$ 
to transform $p^J_\mu$ to $p^{J, rest}_\mu \equiv (m^{jet}_{inv}, 0, 0, 0)$.
\item Boost $p^i_\mu$ by $\vec{\beta}_J$, and obtain $p_\mu^{i, rest}$ .
\item Clustering subjet with $p_\mu^{i, rest}$ and $R_{subjet}$.
\item Sort jets into decreasing $E_i^{rest}$ order and label N most energetic subjets.
\item Get four momenta  of the N subjets in the lab frame, $q^i_\mu$.
\item $\nness =  \frac{2}{ (m^{jet}_{inv})^2}  \sum_{k \in J}  
      \min \{q_1 \cdot p_k,  \dots,  q_N\cdot p_k\}$.
\end{enumerate}
$\diness$ distributions of jets with $ 110 \gev \le m^{jet}_{inv}  \le 130 \gev$, $p_T > 200 \gev$, $|y| < 2.5$
are shown at Fig. \ref{nJetti}.

Applying $\nness$ with $N=2$, we investigate discovery potential of the SM Higgs with $m_H = 120 \gev$,
through the fully hadronic $pp \to W\h, Z\h$ channels 
where both $W/Z$ and $\h$ are highly boosted enough to be identified as single fat jets.
The PYTHIA 6.4.23 \cite{Sjostrand:2006za}  
with the ATLAS MC09 parameter tune \cite{ATL-PHYS-PUB-2010-002} 
and the modified leading-order MRST2007 parton distribution functions \cite{Sherstnev:2007nd}
are used to generate both the signal samples, $pp \to \h V$, and 
the background samples, $pp \to VV$, $Vj$, $tt$, $tV$, $tj$ and $jj$ events at  $14 \tev$.
No K-factors are applied to emulate higher order effects.

FastJet 2.4.2 \cite{Cacciari:2005hq} with the SISCone plugin \cite{Salam:2007xv} is used for the jet clustering,
and the SISCone in spherical coordinates is used for the rest frame subjet clustering.
SISCone jet algorithm has two parameters; the jet radius, $R$, and the overlap threshold, $f$.
For the fat jet tagging using $\diness$, 
the jet radius should be set to maximize the mass peak of the signal,
and the optimal subjet radius should be set to maximize the discrimination power of $\diness$.
With $m_H = 120 \gev$, 
we use $R = 0.8$, overlap threshold $f = 0.75$ for jet clustering,
and $R = 0.6$, $f = 0.75$ for subjet clustering.

The b-tagging, c-jet misidentification and light-jet misidentification probabilities, 
$\epsilon_b$ , $\epsilon_c$ and $\epsilon_{mis}$, are crucial factors to the analysis of this letter.
According to Ref. \cite{ATL-PHYS-PUB-2009-088}, 
$\epsilon_b \sim 70\%$(corresponding to $\approx 50\%$ signal efficiency)
with  $\epsilon_{mis} \sim 1\%$ and $\epsilon_c \sim 10\%$  is expected to be achieved.
Although these values are estimated using C/A MD-F algorithm, 
SISCone jet with the rest frame subjet algorithm can also provide required information;
and, thus, compatible with it.
Therefore, we employ these values for the analysis in this paper.
For estimating the effects on the double b-tag performance of the rest frame subjet,
separate event samples are generated with $B$-mesons are set to be stable;
then, b-tag is assigned to the each subjet with probability $\epsilon_b$ for b-subjet,
$\epsilon_c$ for c-subjet, and $\epsilon_{mis} $ for QCD jet.

To select the signal events which contain two fat jets from the boosted $\h$ and $V$ 
of fully hadronic $pp \to W\h, Z\h$ channels
while reducing QCD backgrounds, we require :
\begin{enumerate}
\label{pt, eta cut}
\item No hard leptons with $p_T > 10 \gev$, except those from B-mesons.
\item Two hardest jets $j_1$ and $j_2$ with $p_{T, j_{1,2}} > 200 \gev$
\item $|y_{j_{1,2}} | < 2.5$  and $|y_{j_1} - y_{j_2}| < 2.0$
	where $y_{j_{i}}$ is rapidity of $j_i$. 
\item Require third hardest jet, $j_3$,  to be not hard : $p_{T, j_3} < 30 \gev$
\item  $b$-cut : both of the two energetic rest frame subjets of the Higgs candidate jet are b-tagged.
\item $\diness$-cut : $\tau_{2, j_1}^{j}, \tau_{2, j_2}^{j} < \nnessCut$
\item \label{symmCut} $\cos \theta_{s}$-cut : in the jet rest frame, angles, $\theta_s$, between two leading rest frame subjets and the boost axis 
should be large enough to be $\cos \theta_s < 0.8$.
\end{enumerate}
To check whether leptons are comes from semi-leptonic decays of $B$-meson, 
we use Pythia's internal data; we have checked that it gives good approximation for the lepton isolation
for the purpose of this paper.

$\cos \theta_{s}$-cut plays similar roles of 
symmetry cut, $y_{cut}$, introduced in Ref. \cite{Butterworth:2008iy}.
To be reconstructed properly, two b-quarks of the Higgs jet should be energetic;
and it results smaller $\cos \theta_{s}$.
$\cos \theta_{s}$-cut  also reduces the effects of UE.
As mentioned previously,
UE particles are concentrated along the boost axis in the jet rest frame.
Since they are mixed with soft particles from the Higgs jet, however, removing these particle is difficult;
and removing them affects the jet mass.
By requiring $\cos \theta_{s}$-cut,
we can select the fat jets which leading two rest frame subjets are not much contaminated by UE,
while preserving the jet mass and reducing QCD backgrounds; 
for SISCone jets, we found that it is a more efficient then removing the soft particles near the boost axis.

We define signal and background events as
which one of  $j_1, j_2$  is the Higgs candidate jet, 
\textit{i.e.} has $m_{inv}^{jet}$  in the range between $120 \pm 10 \gev$ and passes $b$-cut,
and other jet has $m_{inv}^{jet}$ in the range between  $( m_W \pm \mWidth$ or $ m_Z \pm \mWidth) \gev$.
The result before the $b$-cut step of this scheme
is shown at Table \ref{Cross-sections of the signal events}.
The cross section for signal and background events
after applying $b$-cut, $\diness$-cut or both cuts 
are shown at Table \ref{Correlations and cross-sections between two cut}.
Finally, about $36\%$ of QCD dijet backgrounds and 
less than  $5\%$ of the Higgs and other background processes
are filtered out by $\cos \theta_{s}$-cut.
With $\mathcal{L} \sim \lum \ifb$, 
expected $m_{jet}$ distributions are shown at Fig. \ref{mJet};
note that the mass peaks at $m_W$, $m_Z$ should be clearer than the Higgs signals
and we expect they can be used for the calibrations.
The signal / background ratio, S / B, is about 30 / 200; 
and, thus, the statistical significance of the signal is about  $\signalsigma \sigma$.
With a more conservative b-tagging efficiency of $60\%$ and a light-quark jet fake rate of $2\%$,
it is decreased to about $1.5 \sigma$.

\begin{figure}
\includegraphics[width=0.45\textwidth]{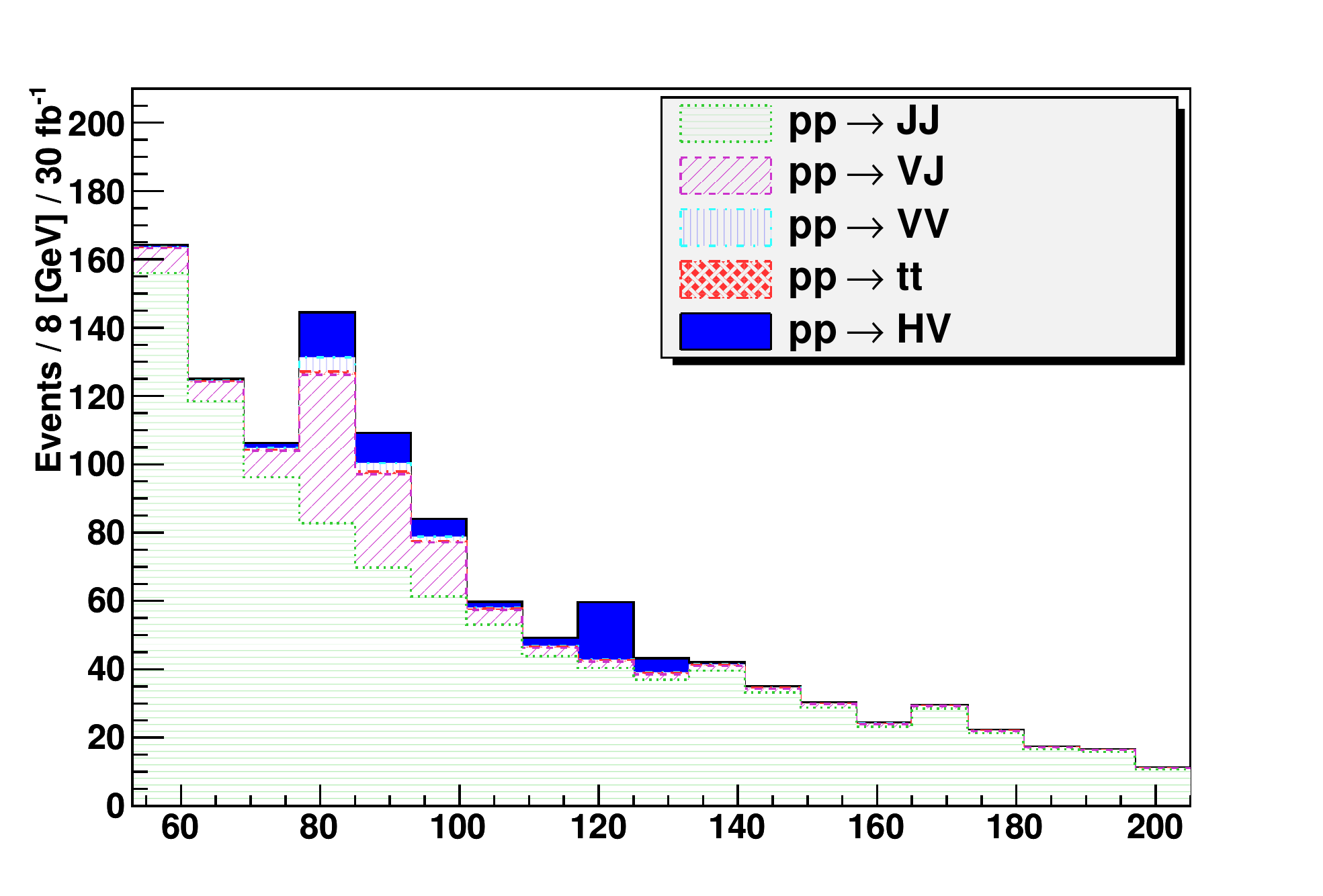}
\caption{Expected $m_{jet}$ distributions of the signal and background events that passes all cuts 
with $\mathcal{L} \sim \lum \ifb$. 
Pythia with the ATLAS MC09 parameter tune and SISCone jet with $R=0.8$ is used.
The $JJ$ sample includes $bb$.
The b-tag efficiency,  c-jet misidentification and light-jet misidentification probabilities  are 
assumed to be $70 \%$, $10 \%$, $1 \%$.}
\label{mJet}
\end{figure}

\begin{table}
\begin{tabular}{|l|c|c|c|}
\hline
$\sigma ( \fb )$     & $p_T$-cut & $y$-cut & $p_T^{j_3}$-cut \\
\hline
$\sigma(pp \to \h V)$ & $2.5\times 10^{1} $& $2.4\times 10^{1} $ & $1.4\times 10^{1} $  \\ 

$\sigma(pp \to VV)$  & $4.1\times 10^1 $  & $3.6\times 10^1 $  & $1.1\times10^{1}$   \\  

$\sigma(pp \to tt)$    & $3.4\times 10^3 $  & $3.0\times 10^3 $ & $1.9\times 10^2 $  \\ 

$\sigma(pp \to Vj)$   & $1.4\times 10^3 $ & $1.3 \times 10^3 $  & $4.2\times10^2$ \\

$\sigma(pp \to gg )$   & $ 1.8\times 10^5 $& $ 1.5\times 10^5 $  & $2.9\times10^{4}$   \\
$\sigma(pp \to qq )$   & $ 1.7\times 10^4 $& $ 1.1\times 10^4 $  & $3.6\times10^{3}$   \\
$\sigma(pp \to qg )$   & $ 1.3\times 10^5 $& $ 1.0\times 10^5 $  & $2.3\times10^{4}$   \\
$\sigma(pp \to bb )$   & $ 4.6\times 10^2 $& $ 4.2\times 10^2 $  & $1.0\times10^{2}$   \\
\hline
\end{tabular}
\caption{Cross section for the signal and background events which
one of  two hardest jet has $m_{inv}^{jet}$  in the range between $120 \pm 10 \gev$ 
and other jet has $m_{inv}^{jet}$ in the range between  $( m_W \pm \mWidth$ or $ m_Z \pm \mWidth) \gev$.
SISCone jet with $R=0.8$ is used.}
\label{Cross-sections of the signal events}
\end{table}

\begin{table}
\begin{tabular}{|l|c|c|c|}
\hline
 $\sigma ( \fb )$  & $b$-cut &$\diness$-cut &  $b$-cut $\&$ $\diness$-cut  \\

\hline
$\sigma(pp \to \h V)$ & $0.4\times 10^{1} $& $0.2\times 10^{1} $ & $0.9\times 10^{0} $  \\ 

$\sigma(pp \to VV)$  & $0.2\times 10^{-1} $  & $0.2\times 10^{-1} $  & $0.1\times10^{-2}$   \\  

$\sigma(pp \to tt)$    & $0.4\times 10^1 $  & $0.7\times 10^1 $ & $0.3\times 10^{-1} $  \\ 

$\sigma(pp \to Vj)$   & $0.5\times 10^1 $ & $0.2 \times 10^2 $  & $0.2\times10^0$  \\

$\sigma(pp \to gg )$   & $ 0.2\times 10^3 $& $ 0.1\times 10^3 $  & $0.3\times10^{1}$   \\
$\sigma(pp \to qq )$   & $ 0.2\times 10^2 $& $ 0.8\times 10^2 $  & $0.4\times10^{0}$   \\
$\sigma(pp \to qg )$   & $ 0.1\times 10^3 $& $ 0.2\times 10^3 $  & $0.2\times10^{1}$   \\
$\sigma(pp \to bb )$   & $ 0.1\times 10^0 $& $ 0.3\times 10^1 $  & $0.4\times10^{-1}$   \\
\hline
\end{tabular}
\caption{Cross section for for the signal and background events 
with a mass window of $20 \gev$ centered on the mass peak
after applying $b$-cut, $\diness$-cut or both cuts. }
\label{Correlations and cross-sections between two cut}
\end{table}

The uncertainties of the statistical significance of signals
are largely come from mass resolution of jet masses,
and modeling of gluon splitting into $b\bar{b}$ 
since most backgrounds come from light quark jets and gluon jets with the gluon splitting.
Both of $\diness$-cut and b-tagging use the leading subjets' informations and, thus, 
there is a loose correlation between them; 
a jet of lower $\diness$ is more likely to pass b-tag cut.
About $2\%$ of light quark jets and $6\%$ of gluon jets
which have $\diness < \nnessCut$  and $m_{inv}^{jet}$ of $120 \pm \mWidth \gev$
is expected to pass $b$-cut.

In conclusion, for the boosted color singlet particle searches,
we have introduced the jet rest frame, the rest frame subjet, and the N-subjettiness.
Using SISCone jet with $\diness$, we see that, 
for the fully hadronic $pp \to \h V$ channels of the SM Higgs with $m_H = 120 \gev$,
the statistical significance of the signals is 
about  $\signalsigma \sigma$ for $14 \tev$ collisions with $\mathcal{L} \sim \lum \ifb$.
Although it will be complementary to the known Higgs search channels,
the scheme suggested in this letter is rather a proof of concept;
the scheme will be improved further
to increase the signal to background ratio, and to make full use of the jet rest frame.
It involves comprehensive studies on theoretical uncertainties of the scheme,
and we left them for the future study.
The rest frame subjet can be defined by any jet algorithms, 
although the effects of UE, and pileup on the scheme depend on the jet algorithm.
We also expect the scheme can also be employed for highly boosted colored particles.

The work of JHK has been supported by the Korean-CERN theory collaboration program 
and KRF-2008-313-C00162.
JHK thanks to Josh Cogan, Hyung Do Kim, and Gavin Salam for useful comments and discussions.


\begin{thebibliography}{99}
\bibitem{Alcaraz:2009jr}
  J.~Alcaraz,
  arXiv:0911.2604 [hep-ex].
  
\bibitem{:2010ar}
    [CDF and D0 Collaboration],
  arXiv:1007.4587 [hep-ex].
  
  
\bibitem{Butterworth:2008iy}
  J.~M.~Butterworth, A.~R.~Davison, M.~Rubin and G.~P.~Salam,
  Phys.\ Rev.\ Lett.\  {\bf 100}, 242001 (2008)
  [arXiv:0802.2470 [hep-ph]].
  
\bibitem{ATL-PHYS-PUB-2009-088}
  [ATLAS Collaboration],
                      ATL-PHYS-PUB-2009-088 (2009)
                      
\bibitem{Plehn:2009rk}
  T.~Plehn, G.~P.~Salam and M.~Spannowsky,
  Phys.\ Rev.\ Lett.\  {\bf 104}, 111801 (2010)
  [arXiv:0910.5472 [hep-ph]].

\bibitem{Ellis:2009me}
  S.~D.~Ellis, C.~K.~Vermilion and J.~R.~Walsh,
  Phys.\ Rev.\  D {\bf 81}, 094023 (2010)
  [arXiv:0912.0033 [hep-ph]].
 
\bibitem{Soper:2010xk}
  D.~E.~Soper and M.~Spannowsky,
  JHEP {\bf 1008}, 029 (2010)
  [arXiv:1005.0417 [hep-ph]].
 
\bibitem{Krohn:2009th}
  D.~Krohn, J.~Thaler and L.~T.~Wang,
  JHEP {\bf 1002}, 084 (2010)
  [arXiv:0912.1342 [hep-ph]].
  

\bibitem{Cacciari:2008gd}
  M.~Cacciari, J.~Rojo, G.~P.~Salam and G.~Soyez,
  JHEP {\bf 0812}, 032 (2008)
  [arXiv:0810.1304 [hep-ph]].

\bibitem{Krohn:2009zg}
  D.~Krohn, J.~Thaler and L.~T.~Wang,
  JHEP {\bf 0906}, 059 (2009)
  [arXiv:0903.0392 [hep-ph]].

\bibitem{Soyez:2010rg}
  G.~Soyez,
  JHEP {\bf 1007}, 075 (2010)
  [arXiv:1006.3634 [hep-ph]].


\bibitem{Almeida:2010pa}
  L.~G.~Almeida, S.~J.~Lee, G.~Perez, G.~Sterman and I.~Sung,
  arXiv:1006.2035 [hep-ph].
  
  
\bibitem{Albrow:2008pn}
  M.~G.~Albrow {\it et al.}  [FP420 R and D Collaboration],
  JINST {\bf 4}, T10001 (2009)
  [arXiv:0806.0302 [hep-ex]].
  
\bibitem{Gallicchio:2010sw}
  J.~Gallicchio and M.~D.~Schwartz,
  Phys.\ Rev.\ Lett.\  {\bf 105}, 022001 (2010)
  [arXiv:1001.5027 [hep-ph]].
 
\bibitem{Salam:2007xv}
  G.~P.~Salam and G.~Soyez,
  JHEP {\bf 0705}, 086 (2007)
  [arXiv:0704.0292 [hep-ph]].

\bibitem{Sapeta:2010uk}
  S.~Sapeta, Q.~C.~Zhang and Q.~C.~Zhang,
  arXiv:1009.1143 [hep-ph].
  
\bibitem{ca_algo}
 Y.~L.~Dokshitzer, G.~D.~Leder, S.~Moretti and B.~R.~Webber,
  JHEP {\bf 9708}, 001 (1997);
 M.~Wobisch and T.~Wengler,
  arXiv:hep-ph/9907280.
  
\bibitem{restTop}
  G.~P.~Salam,
  ``Centre of Mass Top Tagger",
  http://www.lpthe.jussieu.fr/~salam/fastjet/tools.html
     
\bibitem{Stewart:2010tn}
  I.~W.~Stewart, F.~J.~Tackmann and W.~J.~Waalewijn,
  Phys.\ Rev.\ Lett.\  {\bf 105}, 092002 (2010)
  [arXiv:1004.2489 [hep-ph]].
  
\bibitem{Sjostrand:2006za}
  T.~Sjostrand, S.~Mrenna and P.~Z.~Skands,
  JHEP {\bf 0605}, 026 (2006)
  [arXiv:hep-ph/0603175].
  
\bibitem{ATL-PHYS-PUB-2010-002}
  ``ATLAS Monte Carlo tunes for MC09,''
                      ATL-PHYS-PUB-2010-002 (2010)

\bibitem{Sherstnev:2007nd}
  A.~Sherstnev and R.~S.~Thorne,
  Eur.\ Phys.\ J.\  C {\bf 55}, 553 (2008)
  [arXiv:0711.2473 [hep-ph]].
  
\bibitem{Cacciari:2005hq}
  M.~Cacciari and G.~P.~Salam,
  Phys.\ Lett.\  B {\bf 641}, 57 (2006)
  [arXiv:hep-ph/0512210].


\end{thebibliography}
\end{document}
  
\bibitem{Kribs:2009yh}
  G.~D.~Kribs, A.~Martin, T.~S.~Roy and M.~Spannowsky,
  Phys.\ Rev.\  D {\bf 81}, 111501 (2010)
  [arXiv:0912.4731 [hep-ph]].

\bibitem{Falkowski:2010hi}
  A.~Falkowski, D.~Krohn, L.~T.~Wang, J.~Shelton and A.~Thalapillil,
  arXiv:1006.1650 [hep-ph].

\bibitem{Bhatti:2005ai}
  A.~Bhatti {\it et al.},
  Nucl.\ Instrum.\ Meth.\  A {\bf 566}, 375 (2006)
  [arXiv:hep-ex/0510047].
\bibitem{:2010bc}
   {\it et al.}  [ATLAS Collaboration],
  Phys.\ Rev.\ Lett.\  {\bf 105}, 161801 (2010)
  [arXiv:1008.2461 [hep-ex]].
  
\bibitem{Cacciari:2008gp}
  M.~Cacciari, G.~P.~Salam and G.~Soyez,
  JHEP {\bf 0804}, 063 (2008)
  [arXiv:0802.1189 [hep-ph]].

\bibitem{Sjostrand:2007gs}
  T.~Sjostrand, S.~Mrenna and P.~Z.~Skands,
  Comput.\ Phys.\ Commun.\  {\bf 178}, 852 (2008)
  [arXiv:0710.3820 [hep-ph]].

\bibitem{Almeida:2008yp}
  L.~G.~Almeida, S.~J.~Lee, G.~Perez, G.~F.~Sterman, I.~Sung and J.~Virzi,
  Phys.\ Rev.\  D {\bf 79}, 074017 (2009)
  [arXiv:0807.0234 [hep-ph]].

\bibitem{Ellis:2010rw}
  S.~D.~Ellis, C.~K.~Vermilion, J.~R.~Walsh, A.~Hornig and C.~Lee,
  arXiv:1001.0014 [hep-ph].

$\nness$  may resemble the sphericity.
In some cases, actually, both gives similar results. 
The sphericity of the event is defined as
$S = \frac{3}{2} (\lambda_2 + \lambda_3)$
where $\lambda_1 \ge \lambda_2 \ge \lambda_3$ 
are eigenvalues of the standard sphericity tensor $S^{ab} = (\sum_i p_i^a p_i^b) / (\sum_i p_i^2) $ 
and $p_i$ is three momenta of final state particles.
By definition, $0 \le S \le 1$ 
and a dijet event corresponds to $S \approx 0$ and 
an isotropic event to $S \approx 1$.
For a given jet J, the sphericity of J, $S_{rest}^{J}$, can be defined at the rest frame of J
by using $p_i$ in the rest frame of J
and summing constituent particles of J only when calculating $S^{ab}$,
\textit{i.e}.  changing summation $\sum_i$ to $\sum_{i \in J}$ of the definition of $S^{ab}$ .
In an ideal case, with a perfect detector resolution and with no underlying event(UE),
$S_{rest}^{jet}$ and $B_2$ have similar behavior for the boosted Higgs.

However, using $S_{rest}^{jet}$ has several limitations.
Consider a boosted Higgs jet with $\h \to b\bar{b}$, again.
In the rest frame of the jet, 
subjets are broadened as $m_H$ increases and, thus, $S_{rest}^{jet}$  increases.
However, $S_{rest}^{jet}$ can also be raised if there are  another energetic subjets.
$B_2$ can discriminate these two cases by adjusting $R_{subjet}$, but $S_{rest}^{jet}$ can't.
Furthermore, in principle, a color singlet particle may dominantly decays to ${n_p}$ partons.
Even in such cases, $B_{n_p}$ can be used to select a fat jet from the color singlet particle.
But it is hard to differentiate jets containing three subjet from ordinary QCD jet using $S_{rest}^{jet}$.
Another problem is that $S_{rest}^{jet}$ doesn't provide the directional information of subjets.
When axis defined by the b-quark pair is coincide with the boost direction of the Higgs,
a subjet which is directed opposite to the direction of the Higgs
is unlikely to be reconstructed  properly.
$S_{rest}^{jet}$ cannot be used to drop such events and we need to use $B_n$, instead.

In Sec. \ref{sec:The rest frame of a jet and $B_n$}, 
we explain concepts of the jet rest frame and the rest frame subjet.
Then,  N-subjettiness, $\nness$, is introduced and 
different distributions of $\nness$ are shown in Sec. \ref{The N-subjettiness}.
Next, a strategy for the SM Higgs search through fully hadronic channel using $\nness$ 
is described in Sec. \ref{Using $B_2$ for searching Higgs through fully hadronic channel}.
We focus on associated production of the Higgs through  
$pp\to \h(\to b\bar{b}) Z(\to q\bar{q})$ 
where both $\h$ and $Z$ are highly boosted to be identified as a single fat jet.
Finally, we conclude in Sec. \ref{Conclusion.}.

\section{The jet rest frame and the rest frame subjet}\label{sec:The rest frame of a jet and $B_n$}